\documentstyle[preprint,pra,aps,fleqn]{revtex}

\begin{document}
\draft
\title{Negative Binomial States of the Radiation Field and their Excitations are
Nonlinear Coherent States}
\author{Xiao-Guang Wang\thanks{%
email:xyw@aphy.iphy.ac.cn}}
\address{Laboratory of Optical Physics, Institute of Physics,\\
Chinese Academy of Sciences,Beijing,100080,P.R.China}
\author{Hong-Chen Fu}
\address{Institute of Theoretical Physics, Northeast Normal\\
University,Changchun,130024,P.R.China}
\date{\today}
\maketitle

\begin{abstract}
We show that the well-known negative binomial states of the radiation field
and their excitations are nonlinear coherent states. Excited nonlinear
coherent state are still nonlinear coherent states with different nonlinear
functions. We finally give exponential form of the nonlinear coherent states
and remark that the binomial states are not nonlinear coherent states.
\end{abstract}

\pacs{PACS numbers:42.50.Dv,03.65.Db,32.80.Pj,42.50.Vk }


Currently there has been much interest in the study of nonlinear coherent
states[1-4]and their superposition[5-6]. The so-called nonlinear coherent
states are defined as the right-hand eigenstates of the product of the boson
annihilation operator $a$ and a nonlinear function of number operator $\hat{N%
}=a^{\dagger }a,$ 
\begin{equation}
f(\hat{N})a|\alpha ,f\rangle =\alpha |\alpha ,f\rangle ,  \label{1}
\end{equation}
where $f(\hat{N})$ is an operator-valued function of the number operator and 
$\alpha $ is a complex eigenvalue. The ordinary coherent states $|\alpha
\rangle $ are recovered for the special choice of $f(\hat{N})=1.$ A class of
nonlinear coherent states can be realized physically as the stationary
states of the center-of-mass motion of a trapped ion[1].These nonlinear
coherent states can exhibit various non-classical features like squeezing
and self-splitting.

In analogy to the defination of the nonlinear coherent state, we define the
multiphoton nonlinear coherent state as

\begin{equation}
g(\hat{N})a^k|\alpha ,g\rangle =\alpha |\alpha ,g\rangle .
\end{equation}
When the positive integer $k=1,$ the multiphoton nonlinear coherent state reduces to
the nonlinear coherent state(one photon).  One certain type of the above
multiphoton nonlinear coherent state is the  $k$-photon coherent state of
Peremolov type[7-10], which is defined as

\begin{equation}
|\alpha ,k\rangle =\exp (\alpha A_k^{\dagger }-\alpha ^{*}A_k)|0\rangle ,
\end{equation}
where

\begin{eqnarray}
A_k^{\dagger } &=&\left( [\hat{N}/k](\hat{N}-k)!/\hat{N}!\right)
^{1/2}a^{\dagger k}, \\
A_k &=&a^k\left( [\hat{N}/k](\hat{N}-k)!/\hat{N}!\right) ^{1/2}=\left( [\hat{%
N}/k+1]\hat{N}!/(\hat{N}+k)!\right) ^{1/2}a^k  \nonumber
\end{eqnarray}
are the so-called Brandt-Greenberg multiphoton operators[11]. In the above
equation the function $[x]$ is defined as the greatest integer less than or
equal to $x.$ The state $|\alpha ,k\rangle $ is the eigenstate of the
operator $A_k,$

\begin{equation}
A_k|\alpha ,k\rangle =\left( [\hat{N}/k+1]\hat{N}!/(\hat{N}+k)!\right)
^{1/2}a^k|\alpha ,k\rangle =\alpha |\alpha ,k\rangle .
\end{equation}
Comparing Eq.(2) and (5), we see that  $k-$photon coherent states of
Peremolov type are  multiphoton nonlinear coherent states with the nonlinear
function $g(\hat{N})=\left( [\hat{N}/k+1]\hat{N}!/(\hat{N}+k)!\right) ^{1/2}.
$

In a recent paper, Sivakumar[12] pointed out that the excited coherent
states[13-15] 
\begin{equation}
|\alpha ,m\rangle =\frac{a^{\dagger m}|\alpha \rangle }{\langle \alpha
|a^ma^{\dagger m}|\alpha \rangle }  \label{2}
\end{equation}
are nonlinear coherent states and gave an explicit form of the corresponding
nonlinear function $f(\hat{N})=1-m/(1+\hat{N})$. Here $m$ is nonnegative
integer. The work brings us to think the following question: if there exist
other quantum states as nonlinear coherent states? The answer is
affirmative. We find that the well-known the negative binomial state[16-21]
are the nonlinear coherent state defined in Eq.(1) . We also show that the
excited nonlinear coherent state is still nonlinear coherent state with
different nonlinear function $f(\hat{N})$. Finally we give the exponential
form of the nonlinear coherent state and remark that the binomial
states[22-27] are not nonlinear coherent states.

Expanding $|\alpha ,f\rangle $ in terms of Fock states $|n\rangle $ 
\begin{equation}
|\alpha ,f\rangle =\sum_{n=0}^\infty C_n|n\rangle ,  \label{3}
\end{equation}
and substituting the expansion into Eq.(1),we get the recursion relation 
\begin{equation}
C_{n+1}=\frac \alpha {{f(n)\sqrt{(n+1)}}}C_n.  \label{4}
\end{equation}
By assuming that the excited coherent state$|\alpha ,m\rangle $ is the
nonlinear coherent state $|\alpha ,f\rangle $, the coefficients[13] 
\begin{equation}
C_n=\frac{\exp (-|\alpha |^2/2)\alpha ^{n-m}\sqrt{n!}}{[L_m(-|\alpha
|^2)m!](n-m)!}  \label{5}
\end{equation}
for $n\geq m$ and zero for $n<m$. $L_m(x)$ is the Laguerre polynomial of
order $m$ defined by 
\begin{equation}
L_m(x)=\sum_{n=0}^m\frac{(-1)^mm!x^n}{(n!)^2(m-n)!}.  \label{6}
\end{equation}

Substituting Eq.(9) into Eq.(8), one can determine the nonlinear function $%
f(n)$ corresponding to the excited coherent state as 
\begin{equation}
f(n)=1-\frac m{1+n}.  \label{7}
\end{equation}
By a different approach we have obtained the nonlinear operator-valued
function $f(\hat{N})=1-m/(1+\hat{N})$ which is identical to that of
Ref.[12]. Explicitly, the ladder operator form of the excited coherent state
can be written as 
\begin{equation}
\left( 1-\frac m{1+\hat{N}}\right) a|\alpha ,m\rangle =\alpha |\alpha
,m\rangle .  \label{8}
\end{equation}
This shows that the excited coherent state can be viewed as the nonlinear
coherent state.

Supposing that the negative binomial state[20] 
\begin{equation}
|\eta ,M\rangle ^{-}=\sum_{n=0}^\infty (1-\eta )^{M/2}{%
{M+n-1 \choose n}%
}^{1/2}\eta ^{n/2}|n\rangle   \label{12}
\end{equation}
is the nonlinear coherent state and substituting the coefficients of 
Eq.(13) into Eq.(8), we obtain the nonlinear function corresponding to the
negative binomial state as 
\begin{equation}
f(n)=\alpha \eta ^{-1/2}/\sqrt{M+n}.  \label{10}
\end{equation}
Here $\eta $ is a real parameter satisfying $0<\eta <1$ and $M$ is a
non-negative integer. Now the ladder operator form of the negative binomial
state can be written explicitly as 
\begin{equation}
\frac 1{\sqrt{M+\hat{N}}}a|\eta ,M\rangle ^{-}=\eta ^{1/2}|\eta ,M\rangle
^{-}.  \label{13}
\end{equation}

By comparing Eq.(15) with Eq.(1), we conclude that the negative binomial
states are nonlinear coherent states.

Since the excited coherent states are nonlinear coherent states and the
negative binomial states can be reduced to coherent states in certain limit,
we guess that excited negative binomial states are also nonlinear coherent
states. These states can be similarly defined as the excited coherent state.
Next we prove a more general result that excited nonlinear coherent states
are still nonlinear coherent states.

Multiplying both sides of Eq.(1) by $a^{\dagger m}$ from the left yields 
\begin{equation}
a^{\dagger m}f(\hat{N})a|\alpha ,f\rangle =\alpha a^{\dagger m}|\alpha
,f\rangle .  \label{16}
\end{equation}
By the fact 
\begin{equation}
a^{\dagger m}f(\hat{N})=f(\hat{N}-m)a^{\dagger m}  \label{17}
\end{equation}
and the commutation relation 
\begin{equation}
\lbrack a,a^{\dagger m}]=ma^{\dagger (m-1)}  \label{18}
\end{equation}
Eq.(16) can be written as 
\begin{equation}
f(\hat{N}-m)(\hat{N}+1-m)a^{\dagger (m-1)}|\alpha ,f\rangle =\alpha
a^{\dagger m}|\alpha ,f\rangle .  \label{19}
\end{equation}
Using the identity $af(\hat{N})=f(\hat{N}+1)a$ and multiplying the both
sides of the above equation by $a$ from the left, we get 
\begin{equation}
f(\hat{N}+1-m)(\hat{N}+2-m)aa^{\dagger (m-1)}|\alpha ,f\rangle =\alpha (\hat{%
N}+1)a^{\dagger (m-1)}|\alpha ,f\rangle .  \label{20}
\end{equation}
Now we replace $m-1$ by $m$ and multiply the both sides of Eq.(20) from the
left by the operator $1/(\hat{N}+1)$, the following equation is obtained as 
\begin{equation}
f(\hat{N}-m)(1-\frac m{\hat{N}+1})a|\alpha ,f,m\rangle =\alpha |\alpha
,f,m\rangle   \label{21}
\end{equation}
where 
\begin{equation}
|\alpha ,f,m\rangle =\frac{a^{\dagger m}|\alpha ,f\rangle }{\langle \alpha
,f|a^ma^{\dagger m}|\alpha ,f\rangle }  \label{22}
\end{equation}
is the excited nonlinear coherent state. From Eq.(21),we can see that that
the excited nonlinear coherent states are still nonlinear coherent states
with the corresponding nonlinear function $f(\hat{N}-m)(1-\frac m{\hat{N}+1}%
).$ The ladder operator form of the excited coherent states can be obtained
again by setting $f(\hat{N})\equiv 1.$ We have shown that the negative
binomial states are nonlinear coherent states. It is natural that the
excited negative binomial states are nonlinear coherent states based on the
above general result.

The nonlinear coherent state is defined in the ladder operator form. Now we
try to give the exponential form of the nonlinear coherent state.

Eq.(8) directly leads to 
\begin{equation}
C_n=\frac{\alpha ^n}{\sqrt{n!}f(n-1)!}C_0,  \label{23}
\end{equation}
where $f(n)!=f(n)f(n-1)...f(0)$ and $f(-1)!=1.$ Therefore, the nonlinear
coherent state 
\begin{eqnarray}
|\alpha ,f\rangle  &=&C_0\sum_{n=0}^\infty \frac{\alpha ^n}{\sqrt{n!}f(n-1)!}%
|n\rangle   \label{24} \\
&=&C_0\sum_{n=0}^\infty \frac{\alpha ^na^{\dagger n}}{n!f(n-1)!}|0\rangle . 
\nonumber
\end{eqnarray}

One can show that 
\begin{equation}
\lbrack g(\hat{N})a^{\dagger }]^n=a^{\dagger n}g(\hat{N}+n)g(\hat{N}%
+n-1)...g(\hat{N}+1).  \label{25}
\end{equation}
Here $g$ is an arbitrary function of $\hat{N}.$ Then as a key step, using
Eq.(25) with 
\begin{equation}
g(\hat{N})=\frac \alpha {f(\hat{N}-1)},  \label{26}
\end{equation}
the state $|\alpha ,f\rangle $ is finally written in the exponential form 
\begin{eqnarray}
|\alpha ,f\rangle  &=&C_0\sum_{n=0}^\infty \frac{[g(\hat{N})a^{\dagger }]^n}{%
n!}|0\rangle   \nonumber \\
&=&C_0\exp [g(\hat{N})a^{\dagger }]|0\rangle   \nonumber \\
&=&C_0\exp [\frac \alpha {f(\hat{N}-1)}a^{\dagger }]|0\rangle .
\end{eqnarray}
Here, $C_0$ can be determined as 
\begin{equation}
C_0=\left\{ \sum_{n=0}^\infty \frac{|\alpha |^{2n}}{n![f(n-1)!]^2}\right\}
^{-1/2}.  \label{28}
\end{equation}

In fact, by direct verification, we have 
\begin{equation}
\lbrack f(\hat{N})a,\frac 1{f(\hat{N}-1)}a^{\dagger }]=1.  \label{29}
\end{equation}
If the nonlinear coherent states are defined in the exponential
form(Eq.(27)), one can directly obtain the ladder operator form of the
nonlinear coherent state(Eq.(1)) from Eq.(29$).$ In order to confirm the
above result further, we try to get the exponential form of the negative
binomial states from Eq.(27).

By comparing Eq.(1) and Eq.(15) and using Eq.(27),the negative binomial
states can be expressed as 
\begin{equation}
|\eta ,M\rangle ^{-}=\exp \left[ \eta ^{1/2}K_{\dagger }\right] |0\rangle 
\label{31}
\end{equation}
up to a normalization constant. Here 
\begin{equation}
K_0=\hat{N}+M/2,K_{\dagger }=a^{\dagger }\sqrt{M+\hat{N}},K_{-}=\sqrt{M+\hat{%
N}}a  \label{32}
\end{equation}
are the Holstein-Primakoff realization of Lie algebra $SU(1,1).$

The well-known decomposed form of the displacement operator $D_{11}(\xi
)=\exp (\xi K_{\dagger }-\xi ^{*}K_{-})$ is 
\begin{equation}
D_{11}(\xi )=\exp (\eta K_{\dagger })\left( 1-|\eta |^2\right) ^{K_0}\exp
(-\eta ^{*}K_{-}).
\end{equation}
Here $\eta =\frac \xi {|\xi ^{|}}\tanh (|\xi |).$ From the above equation
and Eq.(30),we obtain the displacement operator form of the negative
binomial state as 
\begin{equation}
|\eta ,M\rangle ^{-}=D_{11}(\xi )|0\rangle .
\end{equation}
Here the parameter $\xi $ $=\arctan h(\sqrt{\eta }).$ The displacement
operator form is identical to that obtained in Ref.20.

Next we give a remark that the binomial states are not nonlinear coherent
states. The binomial states are defined as[22] 
\begin{equation}
|\eta ,M\rangle =\sum_{n=0}^M\left[ {%
{M \choose n}%
}\eta ^n(1-\eta )^{M-n}\right] ^{1/2}|n\rangle .  \label{9}
\end{equation}

The ladder operator form of the binomial state was already given by direct
construction[27] 
\begin{equation}
(\sqrt{\eta }J_0+\sqrt{1-\eta }J_{-})|\eta ,M\rangle =\frac{\sqrt{\eta }M}2%
|\eta ,M\rangle ,  \label{14}
\end{equation}
where 
\begin{equation}
J_0=\hat{N}-M/2,J_{\dagger }=a^{\dagger }\sqrt{M-\hat{N}},J_{-}=\sqrt{M-\hat{%
N}}a  \label{15}
\end{equation}
are the Holstein-Primakoff realization of Lie algebra $SU(2)$.

The Eq.(35) can be written in the form 
\begin{equation}
a|\eta ,M\rangle =\left( \frac{{\eta }}{{1-\eta }}\right) ^{1/2}\sqrt{M-\hat{%
N}}|\eta ,M\rangle .  \label{11}
\end{equation}
Note that $\sqrt{M-\hat{N}}|M\rangle =0$, we can not multiply the operator $%
\frac 1{\sqrt{M-\hat{N}}}$ on the both sides of the above equation from the
left and the binomial states are not nonlinear coherent states.

Although the binomial state is not the nonlinear coherent state, it  can
still be written as displacement operator form. The decomposed form of the
displacement operator $D_2(\xi )=\exp (\xi J_{\dagger }-\xi ^{*}J_{-})$ is

\begin{equation}
D_{2}(\xi )=\exp (\eta J_{\dagger })\left( 1+|\eta |^2\right) ^{J_0}\exp
(-\eta ^{*}J_{-}).
\end{equation}
Here $\eta =\frac \xi {|\xi ^{|}}\tan (|\xi |).$ From the above equation and Eq.(36)it
can be directly verified that the binomial state can be written as 
\begin{equation}
|\eta ,M\rangle =D_2(\xi )|0\rangle ,
\end{equation}
where the parameter $\xi $ $=\arctan [\sqrt{\eta /(1-\eta )}].$ This is the
displacement operator form of the binomial state. Eq.(33) and (39) show that
both the negative binomial state and the binomial state can be wriiten as
displacement operator form via the Holstein-Primakoff realization of the
SU(1,1) and SU(2) Lie algebra.

In summary, we have shown that the well-known negative binomial states are
nonlinear coherent states. A more general result, excited nonlinear coherent
states are still nonlinear coherent states, is given in the context. From
this result, we can naturally derive that the excited coherent states and
the excited negative binomial states are nonlinear coherent states. We also
give the exponential form of the nonlinear coherent states and remark that
the binomial states are not nonlinear coherent states. We hope the present
work will stimulate new interest in the study of nonlinear coherent states.

{\bf Acknowledgement:} This work is supported in part by the National
Science Foundation of China with grant number:19875008.

\end{document}